\documentclass[11pt,twoside]{article}
\usepackage{newpasp}
\usepackage{epsf}

\index{Stocke, J. T.}
\index{Shull, J. M.}
\index{Penton, S. V.}
\index{Giroux, M. L.}
\index{McLin, K. M.}
\index{Weymann, R.J.}
\index{Rauch, M.}
\index{Hibbard, J.H.}
\index{van Gorkom, J.C.}

\begin{document}

\title{The Local Ly$\alpha$ Forest: Association of Absorbers with Galaxies, 
    Voids and Superclusters}
\author{John T. Stocke}
\affil{Center for Astrophysics \& Space Astronomy, and Dept. of Astrophysical
\& Planetary Sciences, University of Colorado,
Boulder, CO 80309-0389}

\begin{abstract}
We describe recent discoveries of low column density (N$_{\rm HI}
\leq 10^{14.5}$~cm$^{-2}$) H~I Ly$\alpha$ absorbers made with the 
{\it Hubble Space Telescope} (HST)  which have allowed us a first 
look at gas in local intergalactic space between us and the ``Great Wall''. 
Despite the mere 2.4\,m aperture of HST, these new observations allow 
us to detect absorbers with column densities, N$_{\rm HI} \approx 
10^{12.5}$\,cm$^{-2}$, as low as those found using {\it Keck}/HIRES
at high-$z$.  Owing to the proximity of these absorbers to the Earth, 
the 197 absorbers in our combined GHRS + STIS sample provide our best 
view of the relationship between these absorbers and galaxies, voids, 
and supercluster filaments. Unlike previous results based on galaxy 
surveys near higher-N$_{\rm HI}$ absorbers, we find no evidence that
these lower-N$_{\rm HI}$ absorbers are extended galaxy halos.  Rather, 
the majority (78\%) are associated with large-scale filamentary structures 
of galaxies, while 22\% are found in galaxy ``voids''. Since these Ly$\alpha$ 
absorbers are currently the only baryons detected in the voids, we use 
their properties to estimate that the voids contain 4.5 ($\pm$1.5)\% of    
the Universal baryon density.
\end{abstract}


\section{Introduction}

Unlike virtually all other astronomical objects, Ly$\alpha$ absorbing
``clouds'' were first discovered at great distances ($z\geq$2) due to
cosmological redshifts and the near-UV atmospheric cutoff. It has
only been with the advent of the {\it Hubble Space Telescope} (HST) that 
nearby examples have been found and studied
(Bahcall et~al.\ 1991, 1993; Morris et~al.\ 1991; 
and subsequent HST QSO Absorption Line Key Project papers by Jannuzi 
et~al.  1998 and Weymann et~al. 1998). 
While these absorbers are abundant enough to account for
all baryons at $z\geq$2, their still substantial numbers
at $z\sim$0 imply that $\geq$20\% of all baryons remain in these clouds
locally (e.g., Shull, Penton \& Stocke 1999a; Penton et~al. 2000b;  
Dav\'e et~al. 1998, 2000). Thus,
any account of the present-day distribution of baryons must include an
accurate census of these clouds and the mass associated with them, as
inferred from their column densities and physical extent.  

While the above baryon census is ample reason for studying the local 
Ly$\alpha$ forest in detail, it is also only at low-$z$ that Ly$\alpha$ 
absorber locations can be compared accurately with galaxy locations, so 
that the relationship between these ``clouds'' and galaxies can be 
determined. The degree to which absorbers correlate with galaxies
has been controversial: Lanzetta et~al.\ (1995) and Chen
et~al.\ (1998) argue that the absorbers are the extended halos of
individual bright galaxies (see also Lin et al.\ 2000 and Linder 2000), 
while others (Morris et~al.\ 1993; Stocke et~al.\ 1995; Impey, Petry \&
Flint 1999; Dav\'e et~al.\ 1999; Penton et~al.\ 2001a) argue that the
absorbers are related to galaxies only through their common association
with large-scale gaseous filaments, arising from overdensities in the 
high-redshift Universe.

The intensity of this debate is evident in
the differing views presented at the conference and in these proceedings.
The results reported in this paper come chiefly from an
on-going survey of the local Ly$\alpha$ ``forest'', which utlilizes spectra
like that shown in Figure~1, and which is being conducted by our group at
Colorado (J.M. Shull, S. Penton, M. Giroux \& K. McLin), in collaboration 
with J. van Gorkom (Columbia), C. Carilli
and J. Hibbard (NRAO), and R.J. Weymann and M. Rauch (OCIW).
Current results on the physical conditions and metallicities of the absorbers 
in our sample can be found in the article by J.M. Shull in this volume. 
Since the time of the meeting, we have incorporated 16 additional 
sightlines into this analysis and have discovered
a few more close ($\leq 200 h^{-1}_{70}$ kpc) absorber-galaxy pairs 
in our on-going ground-based redshift survey program (McLin et~al. 2001). 
Neither of these new results significantly affects the conclusions we 
drew at the time of the meeting.
 
\section{The Absorber Sample}

Surprisingly and luckily for local IGM research, the modest HST aperture
is competitive with the 10\,m aperture of the {\it Keck} Telescope ($+$HIRES
spectrograph) in detecting Ly$\alpha$ absorbers because much brighter
targets can be observed. Figure~1 shows an HST$+$STIS (Space Telescope
Imaging Spectrograph) far-UV spectrum of the bright BL Lac Object PKS
2005-489, which detects Ly$\alpha$ absorbers as low in column density, 
N$_{\rm HI}\geq 10^{12.5}$\,cm$^{-2}$, as the best {\it Keck} HIRES data 
(e.g., Hu  et~al. 1995), but within 20,000 km~s$^{-1}$ of the Earth. This 
allows us an unprecendented opportunity to search for faint galaxies 
that could be associated with these absorbers. 

\begin{figure}
\plotone{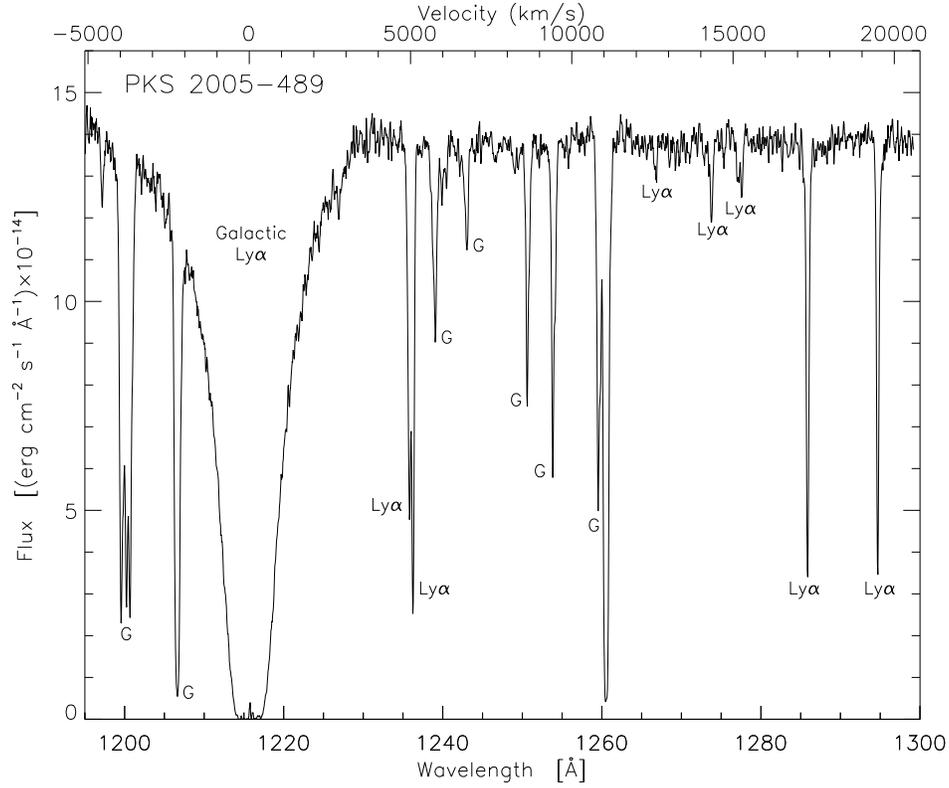} 
\caption{\small
An HST/STIS medium resolution (19 km s$^{-1}$) spectrum (Penton 
et~al. 2001b ) of the bright BL Lac Object PKS 2005-489 illustrates 
the best data obtained for this project. The deep, broad absorption 
at 1216 \AA\ is the damped Ly$\alpha$ absorption due
to the Milky Way. Other Galactic metal lines (S~II, Si~II, Mg~II, 
N~I and Si~III) are marked with a ``G''. The weakest Ly$\alpha$ 
absorbers have column densities, N$_{\rm HI}=10^{12.5}$\,cm$^{-2}$, 
comparable to the weakest absorbers found in the best {\it Keck}/HIRES 
spectra of high-$z$ QSOs.  The heliocentric velocity scale 
at top is for the Ly$\alpha$ absorbers only.}
\end{figure}

All results reported here are based on data toward 15 targets
studied with the Goddard High Resolution Spectrograph (GHRS)
(Penton et~al. 2000a,b, 2001a) and 16 targets observed using STIS 
(Penton et~al. 2001b). In the entire survey, we have detected 197 
Ly$\alpha$ absorbers at $\geq 4\sigma$ significance
over a total unobscured pathlength $\Delta z =1.1$.  This
yields $dN/dz\sim$200 per unit redshift at N$_{\rm HI}\geq$10$^{13}$\,cm$^{-2}$ 
or one ``cloud'' every 20\,$h^{-1}_{75}$ Mpc.  The 20\% baryon fraction quoted
above uses this line density, a 100\,$h^{-1}_{75}$ kpc spherical cloud
extent and the standard 10$^{-23}$\,ergs\,s$^{-1}$\, cm$^{-2}$\,Hz$^{-1}$\,
sr$^{-1}$ local ionizing flux value (Shull et~al. 1999b). To ensure that 
absorber-galaxy statistics are not biased due to incomplete galaxy survey 
information, we will quote results only for a subset of 86 absorbers that 
lie in regions of space surveyed for galaxies down to at least $L^*$ 
based upon the CfA redshift survey catalog (Huchra et al.\ 1992), February 8, 2000 
version and/or our own galaxy redshift survey work (McLin et~al. 2001).  
Preliminary results for a smaller group of 56 absorbers in regions surveyed 
down to 0.1$L^*$ also are presented. Most of the results presented herein 
are described in detail for the GHRS sample in Penton et~al. (2001a). 

\section{Absorber Galaxy Relationships} 

The two-point correlation function (TPCF), which measures
the clustering of local Ly$\alpha$ absorbers, is similar to that found at
high-$z$ in that there is a 4$\sigma$ excess power over random at
$cz\leq$200 km s$^{-1}$, but with no excess power at any larger $\Delta (cz)$. 
Impey, Petry \& Flint (1999) found a similar
result using lower resolution spectra. The absence of significant
clustering of these absorbers is strong evidence that Ly$\alpha$ clouds do
not arise in galaxy halos. But some investigators (see e.g., Fernandez-Soto 
et~al. 1996) suggest that unresolved blends in Ly$\alpha$ absorption lines 
could cause us to underestimate the clustering of Ly$\alpha$ absorbers. 
However, the local galaxy halo TPCF has a large amount of excess power out 
to $\sim$1000 km s$^{-1}$ as well as a signature of galaxy voids, neither of 
which are present in the Ly$\alpha$ TPCF (Penton et~al. 2001a). Also, 
observations with STIS using the echelle gratings as well as
Ly$\beta$ observations with FUSE (Shull et~al. 2000) 
strongly suggest that there are at most 3 blended lines in each of our 
detected Ly$\alpha$ absorbers.  Additionally, our absorber sample is at 
such low-$z$, that we can view the relationship
between absorbers and galaxies directly, so that the TPCF is not as 
essential to understanding galaxy-absorber relationships as it is at high-$z$.
 
Using available bright galaxy redshift survey results and our own redshift 
survey work, we have searched for the nearest known galaxies to these 
absorbers and have found only a few close matches among the subset of 86 
absorbers in sky regions surveyed down to at least $L^*$, although the 
detected close galaxies are almost always well below
$L^*$ in brightness. For example, 6 of the 8 galaxies within 
$175h^{-1}_{70}$ kpc of an absorber are
more than 2 magnitudes fainter than $L^*$ and almost all of these are 
in the Virgo Cluster (6 of these same 8), where the galaxy density is known 
to be much higher than elsewhere (Impey, Petry \& Flint 1999). For 11 Virgo 
cluster absorptions, Impey et~al. had found that the nearest-neighbor 
distances were so large that it was not possible to identify the absorber as
a single galaxy halo. We find the same result: the median nearest galaxy 
neighbor distance from a Ly$\alpha$ absorber in the $L^*$ survey region  
sample is $800h^{-1}$ kpc, while the median nearest galaxy neighbor to another  
$L^*$ galaxy is $200h^{-1}$ kpc away in the same region (see Figure 2). 
Thus, even for galaxies as bright as $L^*$ it is problematical to
identify a low column density absorber with a single galaxy. This and other 
features of the nearest neighbor distributions are shown in Figure 2, 
where the Ly$\alpha$ absorber sample has been split into two
equal parts by number at 86 m\AA. While the stronger absorbers follow 
the galaxy nearest-neighbor distances more closely than the weaker absorbers 
(the two Ly$\alpha$ subsamples differ at the 4.5$\sigma$ level), the 
stronger absorbers cluster much more weakly with galaxies than
galaxies cluster with each other (5$\sigma$ difference) but the weaker 
absorbers are not randomly distributed relative to galaxies either 
(7$\sigma$ difference). Thus, even with a much
larger and more uniformly surveyed sample, our current conclusions do not 
differ from our earlier work (Stocke et~al. 1995).

\begin{figure}
\plotone{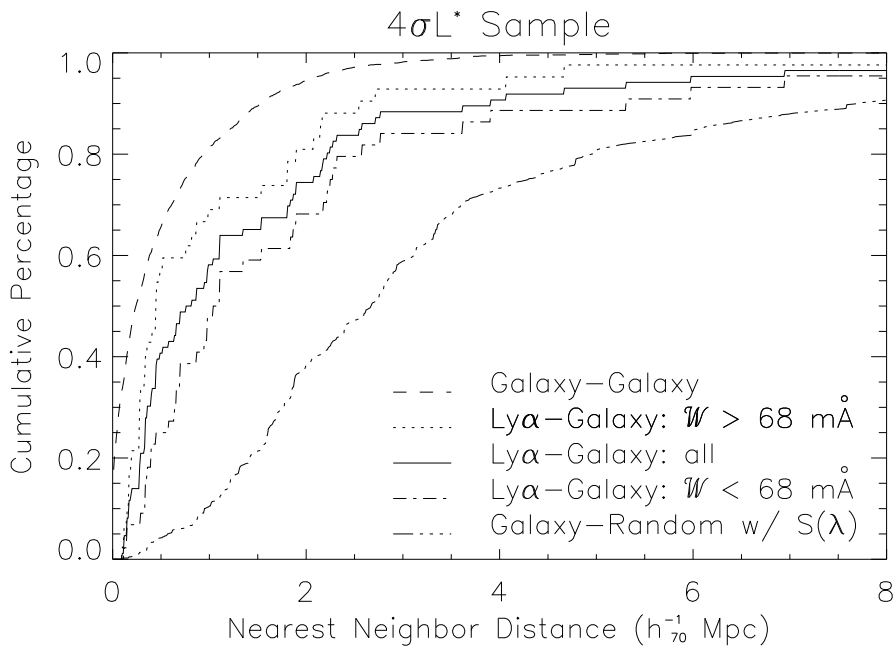}
\caption{\small
Nearest-neighbor cumulative distribution functions (CDFs) for the 86
absorbers in our $4\sigma L^*$ sample (definite Ly$\alpha$ absorbers
in regions surveyed for galaxy redshifts down to at least $L^*$).
The solid line shows nearest-neighbor distances for the full $4\sigma L^*$
sample; the dotted and dash-dotted lines are for the upper half and
lower half subsamples respectively. The dashed line is the nearest-neighbor
distribution for galaxies within the same survey volume, while the triple
dot-dashed line is the nearest-neighbor distribution for random locations
in the survey volume, corrected for the sensitivity function, S($\lambda$),
of our HST GHRS + STIS observations. Each of these CDFs differ
significantly from each other (see text).   } 
\end{figure}

Furthermore, for those few galaxies found close ($\pm$300 km s$^{-1}$ 
in $\Delta(cz)$ and at
impact parameters $\leq 300h^{-1}$kpc on the sky) to a local 
Ly$\alpha$ absorber, the distribution of locations
is random in that volume; i.e., even the galaxies rather near to 
Ly$\alpha$ absorbers do {\bf not} 
concentrate close to the absorber, as would be expected if they were 
physically associated. This is true both for the
$L^*$ subsample (Penton et~al. 2001a) and for the 0.1$L^*$ sample 
(McLin et~al. 2001) and
suggests that even when there is a galaxy close to the absorber, 
the placement of the galaxy is random with respect to the absorber, 
as would be expected if there were no direct physical
association between the two. 
  
Sixteen percent (14 absorbers) 
of these absorbers lie in well-defined galaxy voids, with no known
galaxies within 2-5\,$h^{-1}_{70}$\,Mpc. Deep optical and HI 21\,cm
observations still in progress have failed to locate any galaxies close to
four of these ``void'' absorbers (McLin et~al. 2001; Hibbard et~al. 2001) 
down to impressively low limits due to their proximity 
to the Earth ($cz$=2000-3000 km s$^{-1}$). For the four nearest 
``void absorbers'', those in the MRK 421, MRK 509, VII Zw 118 and 
HE1029-140 sightlines, the CfA redshift survey catalog finds no
$L\geq0.1L^*$ galaxies within 2-5\,$h^{-1}_{70}$\,Mpc, while pencil-beam 
optical spectroscopy finds no very faint ($M_B\leq -13$ to $-13.5$) galaxies 
within 100-250 $h^{-1}_{70}$\,kpc of the
absorber and HI observations with the VLA finds no H I emitting object 
with $M\geq10^8M_{\sun}$ present in a much larger region around three 
of these four absorbers (HE1029-140 has not yet been observed in H~I 
21 cm emission).  Thus, the absence of galaxies in voids and 
near ``void absorbers'' is confirmed and the only
detectable baryons in voids are the Ly$\alpha$ absorbers. By correcting 
for the somewhat variable sensitivity of our Ly$\alpha$ survey with 
redshift, we find that 22\% of all absorbers are in voids
and thus (recalling that the full Ly$\alpha$ absorber density accounts 
for $\sim20\%$ of all baryons locally), the baryon density in voids is 
only 4.5 ($\pm$1.5)\% of the total mean baryon density of the Universe 
determined by [D/H] ratio; see Penton et~al.\ (2000b, 2001a) for the 
detailed calculations of $\Omega_{\rm baryon}$ in Ly$\alpha$ absorbers. 
If Ly$\alpha$ absorbers have no bias factor, then the 
supercluster-to-void density ratio is at least 20:1.

The plots of the cumulative distribution of absorber nearest-neighbor 
galaxy distances and of equivalent width (EW) versus impact parameter 
($\rho$) for our full sample are similar to those published previously 
(Stocke et~al.\ 1995) and to the results found by Tripp, Lu \& Savage 
(1998) and Impey, Petry \& Flint (1999). The correlation between EW and $\rho$ 
contains all the salient features (e.g., lack of correlation
at low-N$_{\rm HI}$; we find no correlation whatsoever for our enlarged sample)  
expected from the N-body+hydrodynamic simulations of Dav\'e
et~al. (1999). Dav\'e et~al. interpret this plot as due to large-scale
structure filaments; i.e., the correlation between EW and $\rho$ does {\bf not} 
require either a physical or a causal association 
with individual galaxy halos as proposed by Lanzetta et~al. (1995) and Chen
et~al. (1998). Our TPCF results and discovery of a substantial fraction of
all absorbers in voids supports the Dav\'e et~al. (1999) interpretation. 
Thus, it seems that this plot cannot be used to support the hypothesis 
that all Ly$\alpha$
absorbers are very extended galaxy halos, although we emphasize that 
our absorber sample and that of Lanzetta et~al. (1995) and Chen et~al 
(1998) do not overlap significantly in equivalent width.

In one case, the sightline pair 3C273/Q1230$+$011, separated by
0.91$^\circ$ on the sky, we have a preliminary indication that both the 7
absorbers and 9 known galaxies in this vicinity are aligned along a
single ($>500\,h^{-1}_{75}$\,kpc), elongated ($>$3:1) filament at
$cz$=1000$-$2000\,km\,s$^{-1}$ (Penton et~al. 2001a). Additionally, 
Penton et~al. (2001a) find a strong indication that this particular 
set of absorbers is not unique in this respect; i.e., 
for our full absorber sample there is a strong (4-12$\sigma$) statistical 
alignment of absorbers along galaxy 
filamentary structures. This alignment with filaments is in contrast to 
the absence of any alignment between
absorber locations and galaxy major axes found by Lanzetta et~al. (1995).
Eventually, perhaps with the Cosmic Origins Spectrograph (COS)
on HST, we will be able to use Ly$\alpha$ absorbers and galaxy survey data
(e.g.,  Sloan Digital Sky Survey) to map out the full extent of 
large-scale structure filaments in the local Universe. 

For one case, a close grouping of Ly$\alpha$ absorbers at  $cz\approx$
17,000\,km\,s$^{-1}$ in the
direction of PKS 2155-304, we have good metallicity limits (Shull et~al. 1998)
for low column
density absorbers far from galaxies. 
Deep optical galaxy survey work (McLin et~al. 2001) has
failed to find fainter galaxies closer to the absorbers than the H~I
emitters found previously, the closest of which is $\sim400h^{-1}$ kpc 
away on the sky.  No metal
lines (C~IV and Si~III)  have been detected as yet in the several strong
Ly$\alpha$ systems at this redshift, placing upper limits on the
metallicity of these clouds of $\la$1\% solar.
This result is still preliminary, awaiting better H~I column density
information from new HST (STIS) and FUSE spectra. 
So, while extensive metallicity surveys of Ly$\alpha$ absorbers have not yet been
conducted with HST, it is possible that some clouds (e.g., the ``void absorbers'') 
are devoid of metals, setting significant
``fossil'' constraints on the spread of metals in the early Universe 
due to supernovae winds and galaxy interactions (see e.g.
Hui \& Gnedin, 1997). However, we 
have found at least one absorber that is undetected in C~IV, C~III and 
C~II but has strong O~VI absorption (Tripp et~al., 2001), so that any 
metallicity result based only on lower ionization species alone must be 
taken with caution. The existence of this one Ly$\alpha$ $+$ O VI absorber 
is additional evidence for shock-heated ``clouds'' (see Tripp
contribution to this volume). However, we caution that virtually all 
O~VI absorbers (i.e., Chen \& Ostriker's ``hot-warm'' phase) should be
detectable in Ly$\alpha$ and so should be present in our GHRS+STIS survey;
i.e., it is important not to ``double count'' absorbers when determining
the total baryon content of the local Ly$\alpha$ ``forest''! 

\section{Why All the ``Fuss''?}

For the uninitiated viewing the debate on the relationship between 
Ly$\alpha$ absorbers and galaxies at the Pasadena Conference, one must 
ask whether the issue is primarily semantics or science.
On the one hand, at this conference Ken Lanzetta and Hsiao-Wen Chen 
argued that all higher column density absorbers
are individual, spherical galaxy halos with a nearly unity covering 
factor to Ly$\alpha$
absorption out to impact parameters of $\sim 160 h^{-1}$ kpc. 
In this paper I have presented evidence
that the majority of lower column density absorbers are associated, 
not with individual galaxies,
but with large-scale filamentary structures in which both the galaxies 
and the absorbers are imbedded.
This is the view espoused by the cosmic simulators, who find these very 
extensive gaseous structures throughout their simulations at both high- 
and low-$z$.  At the extremes, it is clear that both views are 
correct: there are certainly some higher column density absorbers that 
must be extended galaxy
halos because they are so close to their ``parent'' galaxy 
($\leq 50 h^{-1}$ kpc; e.g. the Mg~II absorbers) as to be bound to it, 
while there are low column density absorbers demonstrably {\bf not} 
associated with galaxies at all because they are in voids. In between, 
the data could argue for either stance
and the differences might be largely semantic, excepting for the issue 
of metallicity. If by galaxy ``halo'' we mean gas that was once within 
the galaxy and then was expelled to large radii, the local Ly$\alpha$ 
absorbers should contain gas with
metallicities of $\geq$ 0.1 Solar. There is preliminary evidence that 
this is the case for
absorbers within $\sim$100 kpc of the nearest bright galaxy (Chen et~al. 
2001; Stocke et~al. in preparation) but not for the bulk of the forest 
which lies at much larger distances from bright
galaxies (see Figure 2). Metallicities for some of these more 
representative clouds are needed to be
more definitive about the origin of gas in the local Ly$\alpha$ absorbers, 
including the ``void absorbers''. If the bulk of the forest is ``primeval'', 
then the observations can
be used to test the simulations; but if, these absorbers are mainly the 
product of latter-day galaxy winds, then detailed ``microphysics'' 
must be included in the simulations to match the
observational record. Focus on the issue of metallicity may allow 
workers in this field to stick to scientific rather than semantical concerns.  

\acknowledgments

We acknowledge the financial support at the University of Colorado of
grants provided through HST GO programs \#6593, \#8182 and \#8125, and NASA
Theory Grant NAG5-7262. I thank all of my collaborators listed in the 
Introduction for their efforts contributing to these results, especially 
Steve Penton and Mike Shull for the Ly$\alpha$ absorber
analysis, Ray Weymann for the faint galaxy redshift survey
work and John Hibbard and Jacqueline van Gorkom for the H I work at the VLA.

\end{document}